\documentclass[twocolumn,amsfonts,showpacs,superscriptaddress]{revtex4-1}
\usepackage[utf8]{inputenc}

\usepackage{amssymb,amsmath,amsthm,booktabs,mathtools}

\usepackage{amsmath}
\usepackage{amssymb}
\usepackage{esint}
\usepackage{graphicx}
\usepackage[normalem]{ulem}
\usepackage{color}
\usepackage{subfigure}

\begin{document}
\title{Losses in  interacting quantum gases: ultra-violet divergence and its regularization}
\author{Isabelle Bouchoule, Léa Dubois and Léo-Paul Barbier}

\affiliation{Laboratoire Charles Fabry, Institut d'Optique Graduate School, CNRS, Université Paris‐Saclay,\\
91127 Palaiseau cedex, France}
\begin{abstract}
  We investigate the effect of losses on an interacting quantum gas.
  We show that, for gases in dimension higher than one,
  assuming together a vanishing correlation time of the
  reservoir where dissipation occurs, and contact interactions leads to
  a divergence of the energy increase rate.  This divergence is a combined effect of the
  contact interactions, which impart arbitrary large momenta to the atoms,
  and the infinite energy width of the reservoir associated to its vanishing correlation time.
 We show
  how the divergence is regularized when taking into account the finite
  energy width of the reservoir, and, for large energy width, we give
  an expression for the energy increase rate that involves the contact
  parameter. We then consider the specific case of a weakly interacting
  Bose Einstein condensate, that we describe using the Bogoliubov theory.
  Assuming slow losses
  so that the gas is at any time described by a thermal equilibrium, we
  compute the time evolution of the temperature of the gas. 
  Using a Bogoliubov analysis, we also consider the case where the regularization of the
  divergence is due to the finite range of the interaction between atoms. 
\end{abstract}

\maketitle

The effect of the coupling of a many-body quantum system
to an environment attracted a lot of attention in the last years,
in the context of cold atoms experiments.
Engineered coupling was proposed to
realize particular
many-body states~\cite{diehl_quantum_2008,poletti_emergence_2013},
including strongly correlated phases or
highly entangled states~\cite{barreiro_open-system_2011}.
It can also  be used as a resource for
quantum computation~\cite{verstraete_quantum_2009}. 
A particular coupling to an environment, that has received a lot of attention
recently, is realized when the gas suffers
from losses. Losses can produce highly
correlated phases~\cite{roncaglia_pfaffian_2010,kantian_atomic_2009,foss-feig_steady-state_2012,syassen_strong_2008},
induce Zenon effect~\cite{barontini_controlling_2013,nakagawa_exact_2021,garcia-ripoll_dissipation-induced_2009},
drive phase transition~\cite{labouvie_bistability_2016},
lead to non-thermal states~\cite{johnson_long-lived_2017,bouchoule_effect_2020,rossini_strong_2020,bouchoule_breakdown_2021}, and produce cooling~\cite{rauer_cooling_2016,schemmer_cooling_2018,bouchoule_asymptotic_2020}.
In all the works mentioned above, the coupling to the environment
is described assuming  that the 
correlation time of the environment is much smaller than
any characteristic evolution time of the system.
Then 
the time evolution of the system  obeys a universal Lindblad equation
(see the review \cite{daley_quantum_2014})
describing the coupling to an environment of vanishing correlation time.
In this paper we show that this approximation is not always correct.

For a homogeneous single atom loss process, 
the universal Lindblad equation
reads,
for a gas in the continuous space,
\begin{equation}
  \frac{d\rho}{dt}=-(i/\hbar)[H_0,\rho] +\Gamma 
  \int d^{d} {\bf r} \left \{ -\frac{1}{2}\{\psi^+_{\bf r}\psi_{\bf r},\rho \} +
  \psi_{\bf r} \rho \psi^+_{\bf r}
  \right \}
\label{eq:Lindblad}
\end{equation}
where $H_0$ is the Hamiltonian  of the quantum gas, $\rho$ is its
density matrix,
$d$ is the dimension of system, $\psi_r$ annihilates an atom at position
${\bf r}$, and $\Gamma$ is the loss rate.
For simplicity
we consider here a single specy
gas.
Eq.~\eqref{eq:Lindblad} is universal in the sense that
the loss process is 
characterized by a single parameter $\Gamma$, details of the reservoir being irrelevant.

The evolution under the above Lindblad equation is
simple if one assumes the state of the gas
is uncorrelated, for instance within a mean-field approximation : 
the population of each single particle state  decreases exponentially
~\cite{barontini_controlling_2013}.
However, interactions between atoms introduce correlations,
which highly complicates the calculation of the effect of losses.
In cold atoms experiments,
the range of the interaction potential between atoms is typically much smaller
than all length scales in the problem. Then the effect of interactions
is well modeled by a contact interaction term. 
This description of interactions is also a universal model:
details of the interaction potential is irrelevant
and interactions are described by a single parameter, the scattering length.
In this paper, we show that the combination of the two
above universal models 
leads to unphysical predictions in dimensions higher than one:
for a gas with contact interaction evolving under Eq.~\eqref{eq:Lindblad},
the increase rate of the energy diverges.

The divergence of the energy increase rate
originates from the following process. 
The contact interaction in the gas is responsible for
singularities of the many body wavefunction
when two atoms meet~\cite{werner_general_2012-1},
leading, in dimension higher than one,
to a diverging kinetic energy, this divergence
being counterbalanced by the interaction energy such that the total
energy   is finite.
The Lindlab dynamics of Eq.~\eqref{eq:Lindblad}
assumes that 
loss events are instantaneous with respect with the gas dynamics:
within the quantum trajectory description
equivalent to the Lindblad dynamics~\cite{daley_quantum_2014}
a loss event corresponds to the instantaneous action
of the jump operator $\psi_{\bf r}$. 
Thus, just after a loss event occurred, the many-body wavefunction of the
remaining atoms is equal to its value just before the loss event.
This wave function presents a singularity when the position
of an atom approaches the position of the lost atom.
The divergence of the kinetic energy associated with this  singularity,
is no longer counterbalanced by interaction
energy: it amounts to an infinite value of the energy in the system.
Note that the infinitely large increase of the energy is made possible by the
infinite energy available in the reservoir involved in the loss process:
the vanishing correlation time is associated to an infinite energy width.

Several mechanisms could lead to a regularization of the above
divergence. 
First, the finite range of the interaction between the atoms
will  introduce a cutoff that prevents the divergence
of the kinetic energy.
Second,
the reservoir has in practice a finite energy width
which 
limits the maximum  energy a loss event can deposit in the system.
In this paper, we consider both regularizations, with an emphasis on the
effect of the finite reservoir energy-width. 

We first propose a model for the loss mechanism, with a finite
energy width  $E_{\rm res}$. 
Using an analysis of the 2-atoms case, we then
derive the expected value of the energy density increase rate
for a gas with contact interactions, valid for large $E_{ \rm res}$.
We find a general expression, that involves the contact parameter.
Although  our derivation concentrates on the
Bosonic case for simplicity of notations, our results are general.
To compute the evolution of the system beyond this limit of
large $E_{\rm res}$, one needs a many-body model of the system  that
includes correlations between atoms introduced by interactions. We
will concentrate on the
case of a weakly interacting Bose-Einstein condensate and we use the Bogoliubov
description. Within this framework, we compute the evolution of the energy.
Assuming  a loss rate much smaller than the relaxation rate of the gas,
the system can
be described locally at any time by a thermal
equilibrium state.
We compute the expected evolution of the temperature under
the effect of losses.
Within the Bogoliubov treatment, we also consider the case
where 
the regularization comes from the finite range of the interactions.

\paragraph{Model for the loss process. }
We consider a gas made of particles of mass $m$
in  dimension $d=1$ (1D), $d=2$ (2D)  or $d=3$  (3D),
and we use periodic boundary conditions in a
box of size $L^d$.
A homogeneous one-body loss process occurs if, at each point, the atoms
are coupled to a continuum. In a gas confined in 1D or 2D, the frozen dimension(s) could
serve as the continuum, if atoms are coupled to an untrapped state. In 3D,
the loss mechanism could be the de-excitation of the atoms, if
the latter are in a metastable state, in which case the
momentum of the emitted photon
provides the continuum for the loss mechanism.
Here instead
we will consider a simpler, yet equivalent model~\cite{SM}, where the loss mechanism is induced
by a noisy coupling to an untrapped internal state,
the different Fourier components playing the role of the continuum. 
This would correspond
to the effect of a noisy magnetic field for magnetically trapped atoms~\cite{bouchoule_asymptotic_2020}.
More precisely, 
we consider a coupling to the reservoir which writes
\begin{equation}
  V=\int d^d{\bf r} \Omega(t)\psi_{\bf r}b^+_{\bf r} +\, h.c. =
 \sum_{\bf p}  \Omega(t)\Psi_{\bf p}B^+_{\bf p} \, +h.c.,
\label{eq:couplingOmegat}
\end{equation}
where $\psi_{\bf r}$, resp. $b_{\bf r}$,  annihilates an atom of the system,
resp. of the reservoir, at position
${\bf r}$, $\Psi_{\bf p}$, resp. $B_{\bf p}$,  annihilates an atom of the system,
resp. of the reservoir, of momentum
${\bf p}$,  $h.c.$ is the abbreviation of  ``hermitian conjugate'' and
 ${\Omega }(t)$  is a noisy function.
${\bf p}$ takes discrete values whose coordinates are multiple of $2\pi\hbar/L$, and 
 we note
  $p=|{\bf p}|$.
 We define the energy-dependent rate $\Gamma(E)$ from the spectral density of $\Omega(t)$
 according to 
\begin{equation}
  \Gamma(E)=\frac{1}{\hbar^2}
  \int d\tau e^{-iE \tau/\hbar}\langle {\Omega }^*(\tau){\Omega}(0)\rangle
  \label{eq:GammaE}
\end{equation}
and we note $\Gamma_0=\Gamma(0)$.
We assume a Gaussian
correlation function such that $\Gamma(E)=\Gamma_0e^{-E^2/(2E_{\rm res}^2)}$,
where $E_{\rm res}$ is the energy
width of the loss process, corresponding to
 a correlation time  $\hbar/E_{\rm res}$. 
The energy of the state of momentum ${\bf p}$ in the reservoir is
$p^2/(2m)$ where $m$ is the mass of the atoms, up to
a constant term that could be compensated by a shift  in $E$
of $\Gamma(E)$ and that we take equal to zero. 
If there would be a single atom in the system, its loss  rate,
obtained within a Born-Markov approximation~\cite{SM},
would be   $\Gamma_0 $.
The Lindblad equation \eqref{eq:Lindblad} is obtained
by making $E_{\rm res}\rightarrow \infty$  at a fixed
value of $\Gamma_0$: $E_{\rm res}$ then no longer play a role
and the parameter $\Gamma_0$ entirely characterizes the loss process.
However, as shown below, in presence of contact interactions between atoms,
such an approximation
leads to a divergence of the energy increase rate in dimension $d>1$.
In this paper, we consider a finite value for $E_{\rm res}$.

\paragraph{Two atoms case.}
Let us first investigate  the behavior expected for a system
comprising initially 2 atoms. 
In addition to the kinetic energy term, the
Hamiltonian contains a contact interaction term.
We go in the center-of-mass frame so that the total momentum is vanishing
 and we consider a state of energy $E_0$.
The two-atoms 
wave function writes $\varphi({\bf r}_1,{\bf r}_2)=\varphi({\bf r}_1-{\bf r}_2)$,
with $\int\int d^d{\bf r}_1d^d{\bf r}_2 |\varphi({\bf r}_1-{\bf r}_2)|^2=1$.
For simplicity of notation, we will consider identical Bosonic atoms and use second
quantization representation, such that
this state reads
$|\varphi\rangle = (1/2)\sum_{\bf p} \varphi({\bf p}) \Psi^+_{\bf p} \Psi^+_{\bf -p}| 0 \rangle$,
where, for ${\bf p}\neq {\bf 0}$,
$\varphi({\bf p})=\sqrt{2}\int d^d{\bf r} e^{i{\bf p.r}/\hbar}\varphi({\bf r})$.
The contact interaction imposes the short distance behavior
$\varphi(r)\simeq u_0(|r|-a_{1D})$ in 1D, 
$\varphi({\bf r})\simeq u_0\ln(|{\bf r}|)/a_{2D})$ in 2D and
$\varphi({\bf r})\simeq u_0(1/|{\bf r}|-1/a_{3D})$ in 3D,
where the parameter $u_0$ depends on $d$, $E_0$ and $L$.
In momentum space, this asymptotic form leads to
the large $p$ behavior~\cite{werner_general_2012-1,werner_general_2012}
\begin{equation}
|\varphi({\bf p})|^2\underset{|{\bf p}|\rightarrow \infty}{\simeq} {\alpha_d}\frac{\hbar^4 |u_0|^2}{p^4},
\end{equation}
where $\alpha_d=8$ in 1D,  $\alpha_d=8\pi^2$ in 2D and $32\pi^2$ in 3D.

The 2-atoms state $|\varphi\rangle$  is coupled by $V$ to the
states
$|{\bf p}\rangle=\Psi^+_{\bf p} B^+_{\bf -p}|0\rangle$
whose energy, equal  to the sum of the
kinetic energies of  the lost atom and the remaining atom,
is $p^2/m $.
For weak enough $\Gamma_0$, one can use 
Born-Markov approximation to compute the
loss rate towards the state $|{\bf p}\rangle$~\cite{SM}. 
Using $\langle {\bf p}|V|\varphi\rangle=\Omega(t)\varphi({\bf p})$,
one finds a loss rate
\begin{equation}
  \gamma(p)=|\varphi({\bf p})|^2\Gamma(p^2/m-E_0).
  \label{eq:gammap}
\end{equation}
We can then compute the initial rate of change of the energy for the trapped atoms:
$dE/dt=-\Gamma E_0+
\sum_{\bf p} p^2/(2m) |\varphi({\bf p})|^2 \Gamma(p^2/m-E_0)$.
Here $\Gamma=\sum_{\bf p} \gamma({\bf p})$ is the total loss rate.
We will assume  that $E_{\rm res}$ is large enough so that there exists a momentum
$p_0$ such that 
$E_0 \ll p_0^2/m\ll E_{\rm res}$, and  
$|\varphi({\bf p})|^2$ takes its large $p$ asymptotic behavior.
Then, the contribution to
$dE/dt$ of decay processes towards momentum states of the remaining atom
of momentum larger than $p_0$ is 
\begin{equation}
  \left . dE/dt\right )_{|{\bf p}|>p_0}=\Gamma_0 \frac{\alpha_d \hbar^4 L^d|u_0|^2}{(2\pi\hbar)^d m}{\cal B},
  \label{eq:dEdt2atoms}
\end{equation}
where
\begin{equation}
  {\cal B}{=} \left \{
  \begin{array}{ll}
\!\!\int_{p_0}^\infty dp\Gamma(p^2/m)/(p^2\Gamma_0)\simeq 1/p_0 & \mbox{ in 1D}\\
\!\!    \pi\int_{p_0}^\infty \frac{dp}{p} \Gamma(p^2/m)/\Gamma_0\simeq \frac{\pi}{4} \ln( \sqrt{mE_{\rm res}}/p_0)& \mbox{ in 2D}\\
   \!\! 2\pi\int_{p_0}^\infty dp \Gamma(p^2/m)/\Gamma_0\simeq   \nu \sqrt{mE_{\rm res}}
    &\mbox{ in 3D}.
    \end{array} \right .
\end{equation}
where $\nu = 6.769\dots$.
In 1D, the result does no longer depend on $E_{\rm{res}}$: the energy change rate
has a well defined finite value when $E_{\rm{res}}\rightarrow \infty$.
In the following we consider only gases in dimension $d>1$. 
Then ${\cal B}$ presents a  UV divergence
when $E_{\rm{res}}\rightarrow \infty$, which leads
to the diverging  energy change rate  announced in the introduction. 
The finite value of $E_{\rm res}$ regularizes this divergence. For large
enough $E_{\rm res}$ however, the contribution of the large ${\bf p}$
states dominates  $dE/dt$, and $dE/dt$  is approximately
given by  Eq.\eqref{eq:dEdt2atoms}.
This two-atoms result could have been derived for two different atoms, such as two
different fermions, providing losses affect  both atomic specy in the same way. 

\paragraph{Many-body case: role of the contact.}
The results above can be generalized to many-body systems containing $N$ atoms since, 
for large enough $E_{\rm res}$, the physics
will be dominated by the 2-body physics presented above.
More precisely, one expects the above results
to hold provided one does a sum over the pairs of atoms.
The relevant quantity will be the contact $C$, which
quantifies
the number of pairs in the
gas\cite{tan_energetics_2008,werner_general_2012-1,werner_general_2012}. 
The contact is  defined by the amplitude of the $1/p^4$ tails of the momentum distribution.
More precisely, 
$C=\lim_{p\rightarrow \infty}W(p)p^4$, where
the momentum distribution is normalized to $\int d^d{\bf p} W({\bf p}) = N$. 
In the two atoms case discussed above, $W(p)=|\varphi({\bf p})|^2L^d/(2\pi \hbar)^d$ such that
$C=\hbar^{4-d} \alpha_d|u_0|^2L^d/(2\pi)^d$. 
 Thus, Eq.~\eqref{eq:dEdt2atoms} generalizes to  a many-body system as
  \begin{equation}
  dE/dt=\Gamma_0 \frac{C}{m}{\cal B} 
\label{eq:dEdtContact}
  \end{equation}
  We emphasize the broad applicability of this expression: it
  is valid both in 2D and 3D, and for Fermions or Bosons. 
  Its validity domain is however restricted to very large $E_{\rm res}$.
  To go beyond this approximation, and to estimate its applicability range,
   one should know the details of the many-body physics.   
  In the following we do the calculation in the case of a weakly interacting
  Bose gas described by the Bogoliubov theory.

\paragraph{Exact treatment for a gas described by Bogoliubov}
In this section we suppose the gas is a Bose condensed gas of density $n$.
Beyond-mean field physics is captured, to first approximation,
by the Bogoliubov theory. In this theory, the Hamiltonian
reduces to
\begin{equation}
  H_{BG}=e_0 L^d + \sum_{\bf p\neq {\bf 0}}\epsilon_{p}  a^+_{\bf p} a_{\bf p},
  \label{eq:HBogo}
\end{equation}
  where
  $a^+_{\bf p}$ 
  creates a Bogoliuov excitation of momentum ${\bf p}$ whose 
  energy is $\epsilon_p=\sqrt{p^2/(2m)(p^2/(2m)+2 gn)}$, and
  $e_0$  is the ground state energy density.
  Bogoliubov operators are bosonic operators which fulfill
  $[a_{\bf p}a^+_{\bf p}]=1$,
and they are  related to the atomic operators by the
Bogoliubov transform
\begin{equation}
  \left \{
  \begin{array}{l}
    \Psi_{\bf p}=u_{ p} a_{\bf p} + v_p a_{-{\bf p}}^+\\
    \Psi_{-{\bf p}}^+=v_{p} a_{\bf p} + u_{p} a_{-{\bf p}}^+\\
    \end{array}
  \right .
  \label{eq:bogotransform}
  \end{equation}
where 
$u_{ p}^2 - v_{ p}^2=1$ and $ v_{ p}^2=  ( f_{ p} +f_{ p}^{-1}-2)/4$,
and $f_{p}={ p}^2/(2m\epsilon_{p})$. We set $u_{0}=1$ and
$v_{0}=0$ such that the above equation also holds for ${\bf p}={\bf 0}$.
Note that we use the symmetry breaking  Bogoliubov approach
that does not conserve atom number\footnote{See appendix C of the Supplemental material for  a discussion on this point.}.

Using the Bogoliubov transformation, 
the coupling to the reservoir,  given Eq.~\eqref{eq:couplingOmegat},
reads
  \begin{equation}
  V= \sum_{\bf p}
     a_{\bf p} (u_p \Omega(t) B_{\bf p}^+ + v_p \Omega^*(t) B_{-{\bf p}})
    + \, h.c. 
    \end{equation}

  We compute the master equation describing the  time evolution of
  the density matrix of the system, $\hat \rho$:
  using second order perturbation theory,
  omitting fast oscillating terms, whose effect averages out, and making the Born-Markov
  approximation,
we obtain~\cite{SM}
\begin{equation}
  \frac{d\hat\rho}{dt}=
  \renewcommand{\arraystretch}{2}
  \begin{array}[t]{l}
    -(i/\hbar)[H_0,\hat \rho] \\
    - \sum_{\bf p}\left \{ \Gamma(\frac{p^2}{2m}-gn-\epsilon_p) u_p^2\left ( \frac{1}{2} \{a^+_{\bf p}a_{\bf p},\rho\}  -a_{\bf p} \rho a_{\bf p}^+ \right ) \right . \\
    \left . + \Gamma(\frac{p^2}{2m}-gn+\epsilon_p)v_p^2 \left ( \frac{1}{2} \{a_{\bf p}a_{\bf p}^+,\rho\}  
    - a_{\bf p}^+ \rho a_{\bf p} \right )
    \right \}
    \end{array}
\label{eq:Bogomastereq}
\end{equation}
where the function $\Gamma(E)$ is given in Eq.~\eqref{eq:GammaE}.
For non interacting atoms,  $v_p=0$, $u_p=1$, $gn=0$  and
$\epsilon_p=p^2/(2m)$, such that
the above equation reduces to  Eq.~\eqref{eq:Lindblad}, as expected.
Correlations between atoms introduced by the interactions are
responsible for the anomalous
terms in $v_p^2$.

The first effect of losses is to decrease the density $n$. 
The difference between $dn/dt$ and $-\Gamma_0 n$ is
of the order of  the density of atoms in the
modes of wave vector $p>\sqrt{mE_{\rm{res}}}$.
We assume that  $E_{\rm res}$ is large enough so that we can make the
approximation  $dn/dt \simeq -\Gamma_0 n$.
Let us now investigate the evolution of the energy $E=\langle H_0\rangle$.
We use the Bogoliuobv approximation $H_0\simeq H_{BG}$, where $H_{BG}$ is given
in  Eq.~\eqref{eq:HBogo}, such that
\begin{equation}
  \frac{dE}{dt}= -\Gamma_0 n A 
  +
\sum_{\bf p}\epsilon_{\bf p}
  \left ( \frac{d\langle a^+_{\bf p} a_{\bf p}\rangle}{dt}\right )_{BG}
  \label{eq:dErondedt}
\end{equation}
where $A=L^d de_0/dn+ \sum_{\bf p} \langle a^+_{\bf p} a_{\bf p}\rangle  d\epsilon_{\bf p}/dn $ and
$(d\langle a^+_{\bf p} a_{\bf p}\rangle/dt)_{BG}$ is
the evolution of $\langle a^+_{\bf p} a_{\bf p}\rangle$ within the Bogoliubov approximation.
Inverting the Bogoliubov transform Eq.~\eqref{eq:bogotransform}, we find that
$a^+_{\bf p}$ and $a_{\bf p}$ depend explicitly on time, via the dependence of 
$u_p$ and $v_p$ on $n$. However, we assume that losses
are slow enough so that one has adiabatic following:
$\langle d(a^+_{\bf p} a_{\bf p})/dt\rangle\simeq 0$~\cite{SM}.
Then $(d\langle a^+_{\bf p} a_{\bf p}\rangle/dt)_{BG}$ reduces to
$(d\langle a^+_{\bf p} a_{\bf p}\rangle/dt)_{BG}=Tr( a^+_{\bf p} a_{\bf p}d\hat \rho/dt )$ and
injecting Eq.~\eqref{eq:Bogomastereq}
we obtain 
\begin{equation}
  \renewcommand{\arraystretch}{2}
\begin{array}{ll}
  \left ( \frac{d\langle a^+_{\bf p} a_{\bf p}\rangle}{dt}\right )_{BG}&=
  -\Gamma(\frac{{p}^2}{2m}-gn-\epsilon_{ p})u_{\bf p}^2    \langle a^+_{\bf p} a_{\bf p}\rangle\\
&   +\Gamma(\frac{{ p}^2}{2m}-gn+\epsilon_{ p})v_{ p}^2\left ( 1+\langle a^+_{\bf p} a_{\bf p}\rangle\right ).\\
\end{array}
\label{eq:dnpdt}
\end{equation}
In the case of a reservoir of infinite energy width, for which  $\Gamma(E)=\Gamma_0$
for any $E$, the above equation reduces to
 $ (d\langle a^+_{\bf p} a_{\bf p}\rangle/dt)_{BG}=
  \Gamma_0(-
 \langle a^+_{\bf p} a_{\bf p}\rangle  +v_{p}^2)$.
We recover here the results
derived for 1D Bose gases~\footnote{see the appendix B of~\cite{johnson_long-lived_2017}}.
In particular, since $ v_{p}^2\simeq (mgn)^2/{\rm p}^4$ at large $p$, we
find that $\langle a^+_{p} a_{ p}\rangle$ develops $1/p^4$ tails.
In dimension 1, such tails are responsible for
a failure of Tan's relation~\cite{bouchoule_breakdown_2021}.
In dimension 2 and 3, such tails lead to the unphysical result that
$d E/dt$  diverges.
Proper physical results are obtained in higher dimensions  only
taking into account the finite energy width of the reservoir.
For very large $E_{\rm res}$, $d E/dt$ is dominated by the
second term of the r.h.s of  Eq.~\eqref{eq:dErondedt}, itself dominated by
the large $p$ terms
for
which $v_p^2\simeq (mgn)^2/p^4$,  $\epsilon_p\simeq p^2/(2m) $,
and $\langle a_{\bf p}^+a_{\bf p}\rangle \simeq 0$.
Evaluating the sum, and 
using the fact that the contact within Bogoliubov
theory is $C=L^d(mgn)^2/(2\pi\hbar)^d$, we recover Eq.~\eqref{eq:dEdtContact}.

\begin{figure}
  \centerline{  \begin{tabular}{ll}
      \renewcommand{\arraystretch}{0.1}
      \hspace*{0.5cm}$(a)$ & \hspace*{0.5cm}$(b)$\\[-0pt]
      {\includegraphics[width=0.25\textwidth,viewport=1 1 500 405,clip]{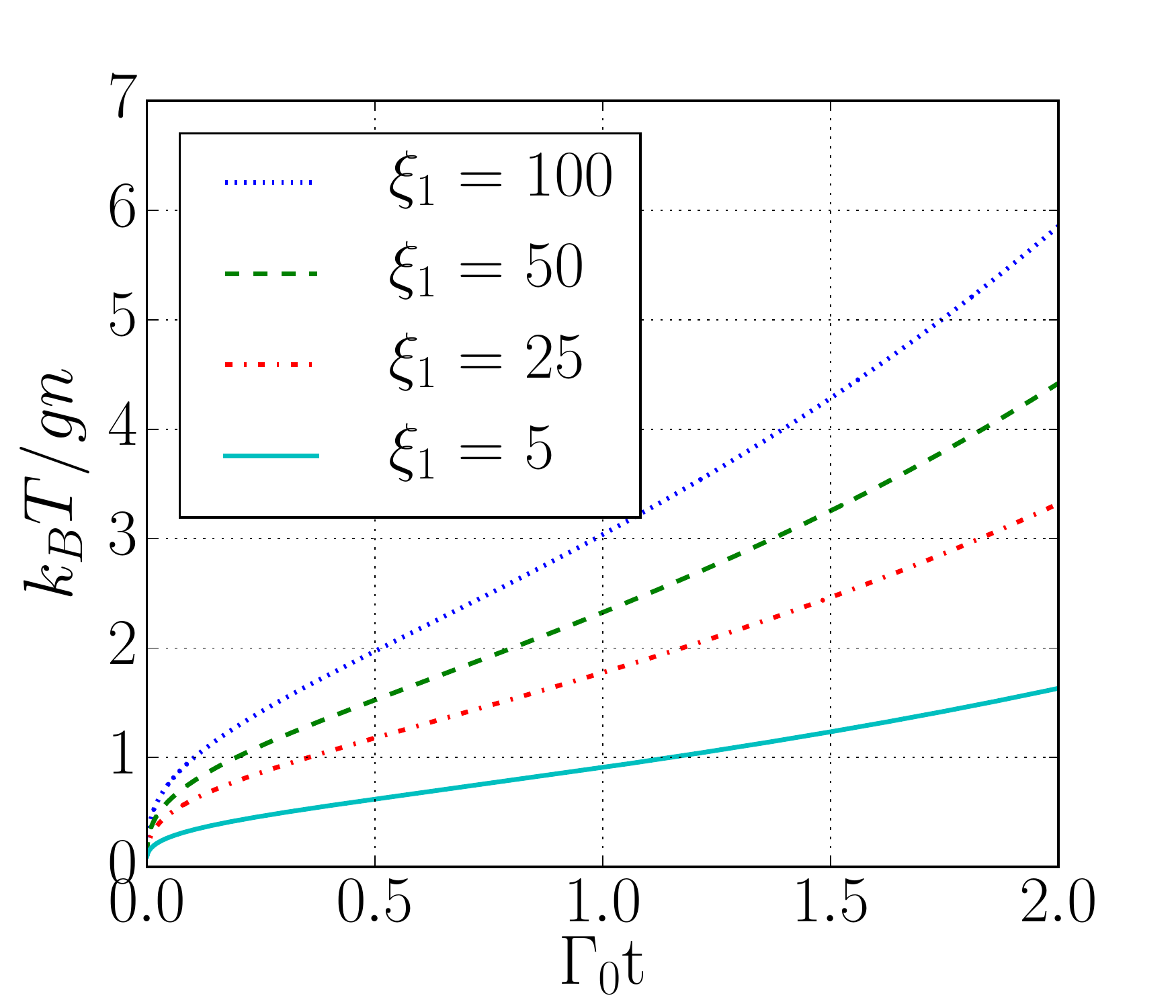}}&
      {\includegraphics[width=0.25\textwidth,viewport=1 1 500 405,clip]{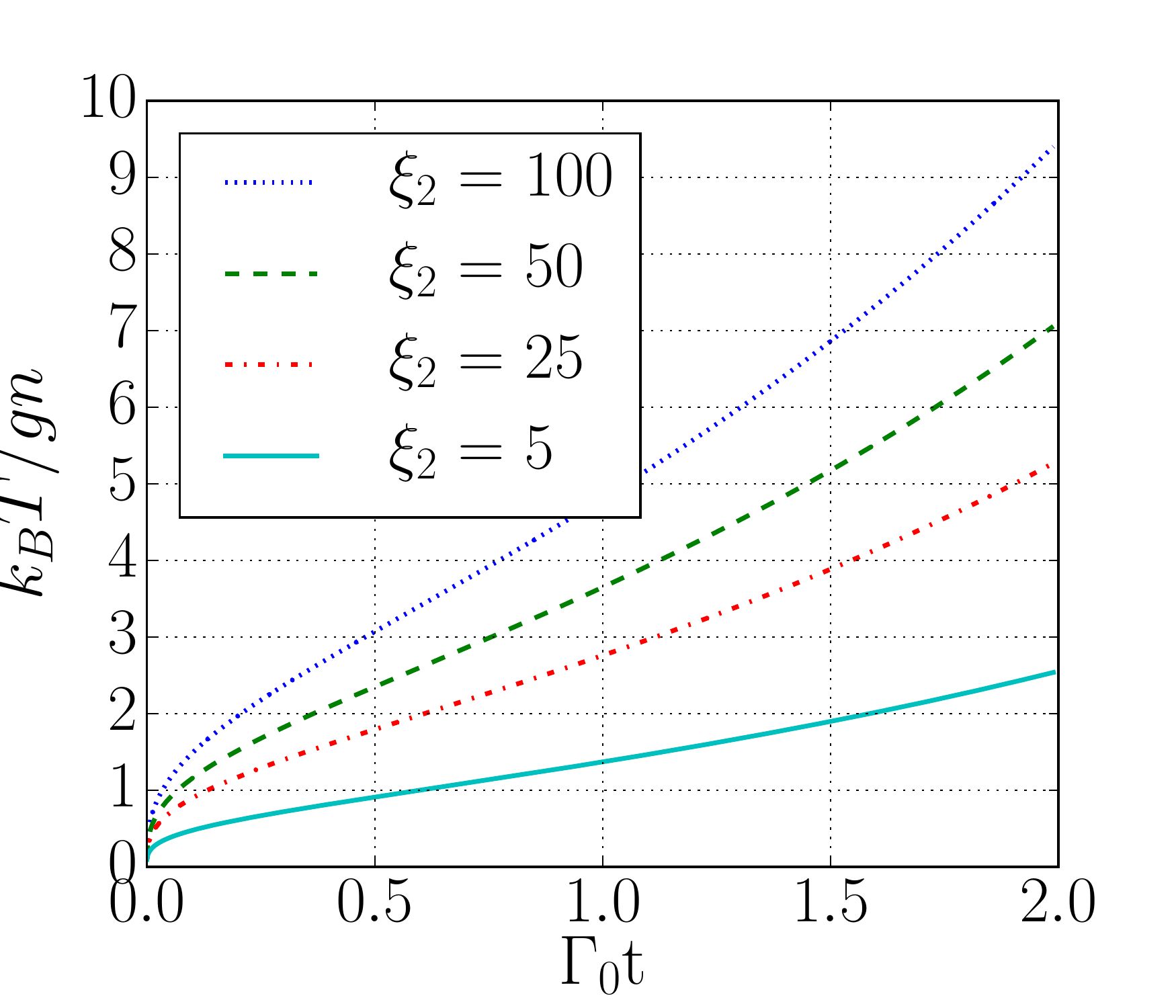}}
  \end{tabular}}
  \caption{Evolution of the temperature of a  3D weakly interacting
    Bose-Einstein condensate under the effect of losses. 
    Temperature is rescaled to the time-dependent chemical potential
    $\mu\simeq gn \simeq gn_0e^{-\Gamma_0 t}$, where $n$ is the atomic density, equal
    to $n_0$ at $t=0$, $g$ the interaction strength  and
    $\Gamma_0$ is the single atom loss rate. The initial
    value is $k_B T(0)/(gn_0)=0.1$.
    In (a) interactions are contact interactions
    but the reservoir has a finite energy
    width $E_{\rm res}$, parameterized by the
    dimensionless parameter $\xi_1=\sqrt{E_{\rm res}/(gn_0)}$. In (b), we assume
    a reservoir of infinite energy width, or equivalently of vanishing correlation time,
    but interactions have a finite range $\sigma$ (see text) and
    $\xi_2=\hbar/(\sigma \sqrt{mgn_0})$.  }
  \label{fig:evolT}
  \end{figure}

The system is ergodic in dimension $d>1$: 
beyond Bogoliubov terms in $H_0$ include couplings between
Bogoliubov modes which, in absence of losses and as long as local observables are concerned,
ensure relaxation towards a thermal state.
Here we assume that $\Gamma_0$ is much smaller than the relaxation rate  so that,
$\hat\rho$ relaxes at any time to the density matrix of a thermal state. The latter is 
characterized by the atomic density $n$ and the energy density,
or equivalently by $n$ and the temperature $T$. 
The energy of the gas fulfills $E=E_{\rm th}(n,T)$,
where $E_{\rm th}(n,T)$ is evaluated injecting the occupation factors
$\langle a_{\bf p}^+ a_{\bf p}\rangle=(e^{\epsilon_p/(k_BT)}-1)^{-1}$ into
Eq.~\eqref{eq:HBogo}.
The time evolution of the gas is entirely characterized by the functions
$n(t)=n_0e^{-\Gamma_0 t}$ and
$T(t)$. 
 In order to compute $T(t)$, we
evaluate $dT/dt$ in the following way. 
Since $E$ is conserved by the thermalization process, the calculation of $dE/dt$
with Eq.~\eqref{eq:dErondedt} and \eqref{eq:dnpdt} is valid, providing one injects
$\langle a_{\bf p}^+ a_{\bf p}\rangle=(e^{\epsilon_p/(k_BT)}-1)^{-1}$ in the r.h.s of
Eq.~\eqref{eq:dnpdt}.
Once $dE/dt$ has been computed 
one can compute $dT/dt$ using
$dE/dt=-\Gamma_0 n (\partial E_{\rm th}/\partial n)_{T}+dT/dt (\partial E_{\rm th}/\partial T)_{n}$.
Calculations are detailed in the appendix. 

In Fig.~\ref{fig:evolT} we present time evolution of the
temperature of the system, for
different values of $E_{\rm res}$.
We find that the ratio $k_B T /(gn)$ is a growing
function of time. This contrasts with the prediction obtained
for phonons, which are the Bogoliubov modes of momentum
$p \ll\sqrt{mgn}$ : in absence of  rethermalization between
Bogoliubov modes, and for $E_{\rm res}\gg gn$,
one expects that, for phonons, $k_B T /(gn)$ takes the
asymptotic value 
$k_BT/(gn)=1$~\cite{grisins_degenerate_2016,johnson_long-lived_2017}.
 The growing of $k_B T/gn$ is due to
the contribution of high-$p$ Bogoliubov modes. 
The growing rate increases with $E_{\rm res}$, as
expected: we expect that $dT/dt$ diverges as $E_{\rm res}$ goes
to infinity.

\paragraph{Regularization by a finite interaction range.}
The Bogoliubov analysis can serve also to describe the regularization of the UV divergence
due to finite interaction range. We consider here a two-body interaction potential
$V(r)=ge^{-r^2/(2\sigma^2)}/((2\pi)^{3/2}\sigma^3)$,
where $ r$ is the distance between the two atoms and $\sigma$ is
the interaction range.
The Bogoliubov transform given in Eq.~\eqref{eq:bogotransform} is still valid, using the
Bogoliubov spectrum~\cite{bogolubov_jr_quantum_2014}
$\epsilon_{\bf p}=\sqrt{p^2/(2m)(p^2/(2m)+2gne^{-p^2\sigma^2/(2\hbar^2)})}$.
One can then compute the effect of losses as above. In the limit of infinite $E_{\rm res}$,
the divergence of $dE/dt$ is regularized by the finite interaction range $\sigma$.
For very small $\sigma$, $dE/dt$ is dominated by the large $p$ term of the sum in
Eq.~\eqref{eq:dErondedt}, for which $v_p^2\simeq (mgn)^2(e^{-p^2\sigma^2/\hbar^2})/p^4$.
Evaluating the sum, we recover Eq.~\eqref{eq:dEdtContact} with
${\cal B}= \pi^{3/2}\hbar/\sigma$. 
As above, from the calculation of $dE/dt$ due to losses,
we compute the time evolution of the temperature
in the system. Fig.~\ref{fig:evolT} shows the time evolution of the temperature, for
different values of $\sigma$.
The evolution of $k_B T/(gn)$ is qualitatively similar to what
is observed for contact interactions but finite energy width of the
reservoir, $\hbar^2/(m\sigma^2)$ playing the role of $E_{\rm res}$.

\paragraph{Conclusion.}
Remarkably, although losses are ubiquitous in experiments, the description
and the understanding of their effect is still at its infancy. 
Before this work,
effect of losses has been studied using the universal Lindblad equation Eq.~\eqref{eq:Lindblad}. 
However,
studies where made either in 1D, in which case 
the divergence of the energy increase rate
does not exists, or for  a gas confined in the lowest band of a lattice,
in which case the lattice period provides a cut-off that prevents
the divergence, or using a mean-field approximation that neglect correlations
between atoms.  
This paper provides the
first prediction for the effect of losses on an interacting quantum
gas in higher dimension and
in the continuum. 
Predictions of this paper could be tested experimentally using
an engineered noisy coupling to an
untrapped state~\cite{bouchoule_asymptotic_2020}
whose energy width $E_{\rm res}$
can be varied.  
This work raises many questions.
How can we extend the results obtained with Bogoliubov to quasi-condensate
describing 2D gases at thermodynamic limit ?
How can the results presented in this paper
be extended to 2-body or 3-body losses ? How can we extend the calculations done in this
paper to other models of quantum gases, such as two-components fermionic gases ?

\paragraph{Acknowledgment}
This work was supported by Palm grant 20P555 and ANR grant ANR-20-CE30-0017-01
The authors thanks D. Petrov and J. Dubail for useful discussions. 

\bibliography{lossesHighD,sm}
\newpage
\appendix
\onecolumngrid
\section{Single-atom case: loss rate}
Let us assume that the system comprises, at $t=0$, a single atom
of momentum ${\bf p}$.
After an evolution time $t$, the state of the system writes
$ c_0 e^{-it p^2/(2\hbar m)} \Psi^+_{\bf p}|0\rangle
+  c_{1}e^{-it p^2/(2\hbar m)} B_{\bf p}^+|0\rangle$,
where $c_0(t=0)=1$ and $c_{1}(t=0)=0$.  
Using perturbation theory, one has
$c_{1}(t)= -i \int_0^t d\tau \Omega(\tau)$ such that 
\begin{equation}
  |c_{1}(t)|^2= \frac{1}{\hbar^2}
  \int_0^t\int_0^t d\tau_1d\tau_2 \Omega^*(\tau_1)\Omega(\tau_2) .
  \label{eq:cp1}
\end{equation}
For weak enough $\Omega$, the evolution of $c_1(t)$ and $c_0(t)$ are slow enough so that
one can consider times $t$ both
much larger than the correlation time of $\Omega$ and 
small enough so that
$c_0$ barely changed and the above perturbation calculation holds.
This is the so called Born-Markov approximation, under which
Eq.~\eqref{eq:cp1} writes
\begin{equation}
  |c_{1}(t)|^2= t \int_{-\infty}^\infty d\tau
\langle \Omega^*(0)\Omega(\tau) \rangle =
t\Gamma_0.
  \label{eq:cp1Gamma0}
\end{equation}
Thus, the loss rate is $\Gamma_0$. It does not depends on the momentum
of the atom. 
The above calculation is similar to the usual Fermi Golden Rule calculation,
the different Fourier components of $\Omega$ playing the role of the different
states of the continuum. 
Note that the fact the energy of the state $B_{\bf p}|0\rangle$ is $p^2/(2m)$
ensures Galilean invariance of the loss process: the loss rate does not dependent
on ${\bf p}$, as expected since one can compute the loss rate
in the moving frame where ${\bf p}={\bf 0}$.

\section{Two-atom case: loss rate towards $|{\bf p}\rangle$}
Here we assume the system initially comprises 2 atoms. 
The loss rate towards a state $|{\bf p}\rangle$, given in Eq.~\eqref{eq:gammap},
is computed using a similar calculation as in the above appendix. 
The initial state $|\varphi\rangle$ is coupled by $V$ to the final states
$|{\bf p}\rangle$ and the state of the system at time $t$ writes
$c_0 e^{-iE_0 t/\hbar} |\varphi\rangle + \sum_{\bf p} c_{\bf p}e^{-it p^2/(\hbar m)} |{\bf p}\rangle$,
where $c_0(t=0)=1$ and $c_{\bf p}(t=0)=0$.
Using perturbation theory, one has
$c_{\bf p}(t)= -i\varphi({\bf p}) \int_0^t d\tau \Omega(\tau)e^{i(p^2/m-E_0)\tau/\hbar}$ such that the population
in the state $|{\bf p}\rangle$ at time $t$ writes
\begin{equation}
  |c_{\bf p}(t)|^2= \frac{|\varphi({\bf p})|^2}{\hbar^2}
  \int_0^t\int_0^t d\tau_1d\tau_2 \Omega^*(\tau_1)\Omega(\tau_2)
  e^{i(p^2/m-E_0)(\tau_2-\tau_1)/\hbar} .
  \label{eq:cp2}
\end{equation}
We then make the Born-Markov approximation as in the above appendix:
for weak enough $\Omega$, the evolution of $c_{\bf p}(t)$ and $c_0(t)$
are slow enough so that
one can consider times $t$ both
much larger than the correlation time of $\Omega$ and 
small enough so that
$c_0$ barely changed and the above perturbation calculation holds.
 Eq.~\eqref{eq:cp2} then transforms into 
\begin{equation}
  |c_{\bf p}(t)|^2= t |\varphi({\bf p})|^2 \int_{-\infty}^\infty d\tau
\langle \Omega^*(0)\Omega(\tau) \rangle 
  e^{i(p^2/m-E_0)\tau/\hbar}.
  \label{eq:cp2BornMarkov}
\end{equation}
We thus  recover Eq.~\eqref{eq:gammap} using $\gamma({\bf p})=|c_{\bf p}(t)|^2/t$.

\section{Noisy time-dependent coupling versus coupling to a  reservoir}
In this paper, for simplicity of notations, instead of considering a  reservoir
where losses occur -- {\it i.e.} a continuum of states--,
we consider a fluctuating time dependent coupling
towards a secondary internal state. This would correspond
for instance 
to a coupling to a different zeeman state 
by a noisy magnetic field, as done in~\cite{bouchoule_asymptotic_2020}.
The different Fourier components of the coupling play the role of the different states
of a  reservoir, as shown below.

In this appendix  we consider a time-independent coupling towards a reservoir. 
The states of the reservoir in which an atom of the system can decay  write
$|{\bf p },i\rangle_{\cal R}=B^+_{{\bf p},i}|0\rangle$: they are labeled by 
their momentum ${\bf p}$ and an additional index $i$ which, in a discretized representation,
labels the states of the continuum.
For instance, atoms in 2D, confined to the ground state of a vertical confining potential,
could be coupled to states untrapped in the
vertical direction and $i$ labels the vertical momentum of the free atoms.
In 3D, in the case of metastable atoms decaying to an untrapped state with the emission
of a photon, $i$ would label the momentum of the emitted photon. 
The coupling between the system and the reservoir
writes
\begin{equation}
  V=\sum_{{\bf p},i} V_{i} \Psi_{\bf p} B^+_{{\bf p},i} + h.c. .
  \label{eq:couplingres}
  \end{equation}
 The fact that the coupling $V_{i}$ does not depend on
${\bf p}$ ensures a homogeneous loss process. 
 The energy of the states $|{\bf p },i\rangle_{\cal R}$ are $E_{{\bf p },i}=p^2/(2m)+E_i$,
 which ensures invariance by Galilean transformation.

 To make the correspondence with the noisy loss model used in the main text,
 let us consider the situation where
 the initial state comprises 2 atoms in the system.
 After an evolution time $t$,
 the population in the states of the reservoir of momentum ${\bf p}$, evaluated using
 second order theory is
 \begin{equation}
P_{\bf p}(t)= |\varphi({\bf p})|^2 \sum_i
  \int_0^t\int_0^t d\tau_1d\tau_2 \frac{|V_{i}|^2}{\hbar^2}
  e^{i(E_i+p^2/m-E_0)(\tau_1-\tau_2)/\hbar} .
  \label{eq:cp2res}
 \end{equation}
 Let us introduce the function
 \begin{equation}
   {\cal R}(\tau)=\sum_i |V_{i}|^2 e^{iE_i\tau}
 \end{equation}
 whose width, denoted $\tau_c$, is called the correlation time of the
 reservoir.
 We assume losses are weak enough so that 
 the typical time of decrease of the initial state population is much
 larger than $\tau_c$. Then, one can consider times $t$ both small
 enough so that the population in the initial state barely changed
 and large enough to fulfill $t \gg \tau_c$. This is the Born-Markov
 approximation, which permits to writes Eq.~\eqref{eq:cp2res} as
  \begin{equation}
    P_{\bf p}(t)= t |\varphi({\bf p})|^2\int_{-\infty}^{\infty} d\tau
   {\cal R}(\tau)e^{i(p^2/m-E_0)\tau}.
  \end{equation} 
  We thus recover Eq.~\eqref{eq:cp2BornMarkov}, providing we
  make the identification
  \begin{equation}
    {\cal R}(\tau) \iff \langle \Omega^*(0)\Omega(\tau)\rangle.
    \label{eq:correspondanceresnoise}
  \end{equation}
  In the frequency domain, the correspondence between the time-dependent
  noisy coupling and the coupling to the reservoir reads
  \begin{equation}
   2\pi n(E) |V(E)|^2/\hbar  \iff \Gamma(-E).
    \label{eq:correspondanceresnoiseE}
  \end{equation}
  where $V(E)$ is the value of $V_i$  for states at an energy $E_i\simeq E$ and
$   n(E)=\sum_i \delta(E_i-E)$
   is the density of state of the reservoir.
  
 

 Here we established 
 the equivalence between a noisy time-dependent coupling and a
 time-independent coupling towards a reservoir  for the short time dynamics.
 For the dynamics to be equivalent at longer times,
 the secondary internal state
 towards which atoms are transferred by the noisy function $\Omega(t)$
 should actually be removed, on a typical time short compared to the
 typical evolution time  of the system.
 This will be the case if
 they are untrapped and quit the zone where the atoms of the
 system are trapped.  If the removal time is much larger than the
 correlation time of $\Omega(t)$, the loss process will be
 correctly captured by the 
 model presented in the main text: the relevant energy width for the loss process
 is the energy width of the function $\Gamma(E)$ given Eq.~(3) of the main text, which is
 nothing else than $\hbar/\tau_c$ where $\tau_c$ is the correlation time of $\Omega$.

 Note finally that the choice of a model of losses based on  a
noisy time-dependent function
is not essential for the results presented in this paper. The master
equation given Eq.~\eqref{eq:Bogomastereq} of the main text,
derived in the appendix below for losses induced by a noisy time-dependent
coupling,  could also be derived
using the coupling to a reservoir given in Eq.~\eqref{eq:couplingres},
at the price of slightly more complicated notations. At the end, one would
recover the same results using the correspondence of
Eq.~\eqref{eq:correspondanceresnoise} and Eq.~\eqref{eq:correspondanceresnoiseE}.

\section{Derivation of the master equation}

Let us first consider  the system that comprises both the gas under study
and the environment in which losses occur.
Its  density matrix is noted $\rho_{\Sigma}$
and its Hamiltonian writes 
$H_{\Sigma} = H_{0} + H_{\mathrm{Res}} + V$, where $H_{0}$ is the Hamiltonian of the quantum gas,
$H_{\mathrm{Res}}$ is the Hamiltonian of the reservoir and $V$ is the coupling Hamiltonian
between the quantum gas and the reservoir.
As explained in the main text, we use the Bogoliubov approximation to describe the quantum gas. 
At time $t$, we assume there is no correlation between
the environment and the system such that $\rho_{\Sigma}(t) = \rho(t) \otimes |0_{R}><0_{R}|$.

We work in the interaction representation, in which an operator $A$ writes
 $\tilde{A}(t) = e^{ i (H_{0} +  H_{\mathrm{Res}}) \frac{t}{\hbar}} A e^{- i (H_{0} +  H_{\mathrm{Res}}) \frac{t}{\hbar}}$, where $A$ is the operator in the Shr\"odinger picture.
The
evolution of the density matrix  $\tilde{\rho}_{\Sigma}$ is given by 
\begin{equation}
    i \hbar \frac{ \mathrm{d} \tilde{\rho}_{\Sigma}}{ \mathrm{d} t}= [\tilde{V}, \tilde{\rho}_{\Sigma}].
    \label{AppC1}
\end{equation}
We consider the evolution of $\tilde{\rho}_{\Sigma}$ on a time $\delta t$ small enough so that
we can restrain 
 to second order perturbation theory.
Integration of the above equation then leads to 
\begin{equation}
    \tilde{\rho}_{\Sigma}(t + \delta t) = \tilde{\rho}_{\Sigma}(t) + \frac{1}{i\hbar} \int_{t}^{t + \delta t} [\tilde{V} (\tau), \tilde{\rho}_{\Sigma}(t)]\mathrm{d}\tau  
     -\frac{1}{\hbar^{2}} \int_{t}^{t + \delta t} \mathrm{d}\tau \int_{t}^{\tau} \mathrm{d}\tau' [\tilde{V} (\tau), [\tilde{V} (\tau'), \tilde{\rho}_{\Sigma}(t)]]
    \label{AppC2}
\end{equation}
The coupling $\tilde{V}(t)$ is obtained from its Shr\"odinger representation given in
Eq. (\ref{eq:couplingOmegat}) of the main text, using the interaction representation of the operators.
The latter are obtained from
\begin{equation}
    \left\{ 
    \begin{array}{lll}
    \tilde{B}_{\mathbf{p}} &= e^{-i \omega_{\mathbf{p}}^{\mathrm{Res}}t} B_{\mathbf{p}} \; & \mbox{with} \; \omega_{\mathbf{p}}^{\mathrm{Res}} = (p^{2}/(2 m)-gn) \hbar \\ 
    \tilde{a}_{\mathbf{p}} &= e^{-i \omega_{\mathbf{p}} t} a_{\mathbf{p}} \;& \mbox{with} \; \omega_{\mathbf{p}} = \epsilon_p/{\hbar}. 
    \end{array}
    \right.
    \label{AppC4}
\end{equation}
where the $\epsilon_p=\sqrt{p^2/(2m)(p^2/(2m)+2gn)}$ is  the energy
of the Bogoliubov mode of momentum ${\bf p}$. The term $gn=\partial E_0/\partial N$ amount to the
energy shift of the ground state energy of the gas when an atom is removed from the system.
It is taken into account by a shift in energy of the reservoir states. 
The validity of this approach appears if one consider
a number-preserving Bogoliubov approach where Bogoliubov operators
conserve atom-number~\cite{castin_low-temperature_1998}.
In this approach, the operator $B_{\bf p}^+$ ( resp. $B_{\bf p}$) in $V$ would be replaced
by the operator $\Lambda B_{\bf p}^+$
(resp. $\Lambda^+ B_{\bf p}p$) where
$\Lambda\simeq (1\sqrt{N})\Psi_{\bf 0}$ remove an atom from the condensate.

We are interested in computing the state of the gas at time $t+\delta t$, irrespective of the
state of the reservoir. We thus trace over the reservoir and compute
$\tilde{\rho} = Tr_{\mathrm{Res}} (\tilde{\rho}_{\Sigma})$.
 Tracing out the reservoir degrees of freedom in Eq.~\eqref{AppC2}, doing the variable change $u=\tau-\tau'$, and using the fact that the reservoir is in its empty state at time $t$, we obtain
\begin{equation}
  \tilde{\rho} (t + \delta t)  = \tilde{\rho}(t) - \frac{1}{\hbar^{2}} \sum_{\mathbf{p}} \int_{t}^{t + \delta t} \mathrm{d} \tau \int_{0}^{\tau} du 
   \left \{ \Omega^*(\tau) \Omega(\tau-u) [
     A_{\mathbf{p}}^{(1)} (u) 
     + A_{\mathbf{p}}^{(2)} (\tau, u)] + h.c. \right \}  
    \label{eq:Ap1Ap2}
\end{equation}
with 
\begin{equation}
\renewcommand{\arraystretch}{1.5}
  A_{\mathbf{p}}^{(1)}(u)  = u_{\mathbf{p}}^{2}e^{i(\omega_{\mathbf{p}} - \omega_{\mathbf{p}}^{\mathrm{Res}}  )u} \left (
    { a}_{\mathbf{p}}^{+} {a}_{\mathbf{p}}  \tilde{\rho}(t) 
    -
    {a}_{\mathbf{p}} \tilde{\rho}(t) {a}_{\mathbf{p}}^{+}\right )
   +  v_{\mathbf{p}}^{2} e^{-i( \omega_{\mathbf{p}}^{\mathrm{Res}}+\omega_{\mathbf{p}} )u} \left (
     {a}_{-\mathbf{p}} {a}_{-\mathbf{p}}^{+} \tilde{\rho}(t)
   -
    {a}_{-\mathbf{p}}^{+} \tilde{\rho}(t) {a}_{-\mathbf{p}}\right ) \\
    \label{AppC7}
\end{equation}
and 
\begin{equation}
  A_{\mathbf{p}}^{(2)}(\tau, u)  = u_{\mathbf{p}} v_{\mathbf{p}}
    [e^{-i 2\omega_{\mathbf{p}}\tau} 
         e^{-i (\omega_{\mathbf{p}}^{\mathrm{Res}}-\omega_{\bf p})u } (
         {a}_{\mathbf{-p}} { a}_{\mathbf{p}}\tilde{\rho}(t)
       -
       {a}_{\mathbf{p}} \tilde{\rho}(t) { a}_{-\mathbf{p}})
      + e^{i 2\omega_{\mathbf{p}}\tau} 
       e^{-i (\omega_{\mathbf{p}}^{\mathrm{Res}}+\omega_{\bf p}) u}
       ({a}_{\mathbf{p}}^{+} {a}_{-\mathbf{p}}^{+} \tilde{\rho}(t) -
       {a}_{\mathbf{-p}}^{+} \tilde{\rho}(t) {a}_{\mathbf{p}}^{+})]
    \label{eq:A2}
\end{equation}
The term $A_{\mathbf{p}}^{(2)}$ oscillates in $\tau$ with the frequency $2\omega_{\bf p}$.
Since losses are weak, we expect that $\tilde{\rho}$ evolves slowly compared to
$\omega_{\bf p}$. Thus, the effect of $A_{\mathbf{p}}^{(2)}$ will average out. In the following we
neglect this term. This constitutes a secular approximation, also called a
rotating wave approximation.

To compute the effect of the term $A_{\mathbf{p}}^{(1)}$, we use a Born-Markov approximation.
This approximation holds assuming the losses are weak enough so that the time evolution of
$\tilde \rho$ occurs on times much larger than the correlation time of $\Omega$, noted $\tau_c$.
Then one can consider a time interval $\delta t$ large enough so that that $\delta t \gg \tau_c$ but
small enough so that the evolution of $\tilde \rho$ is small and the second
order calculation is valid. 
The condition $\delta t \gg \tau_c$ permit to write  Eq.~\eqref{eq:Ap1Ap2} as
\begin{equation}
  \tilde{\rho} (t + \delta t)  = \tilde{\rho}(t) - \frac{\delta t}{\hbar^{2}} \sum_{\mathbf{p}}
  \left (   \int_{0}^{\infty} du \langle \Omega^*(0)\Omega(-u)\rangle A_{\mathbf{p}}^{(1)}(u)
  + h.c.\right ) .
  \label{eq:rhotildedeltatBM}
\end{equation}
Injecting Eq.~\eqref{AppC7} into the above integral, one finds integrals of the form
$\int_0^\infty du \langle \Omega^*(0)\Omega(-u)\rangle e^{-i\nu u}$, where
$\nu=\omega_{\mathbf{p}}^{\mathrm{Res}}\pm\omega_{\bf p}$.
We write them as $\hbar^2\Gamma(\nu)/2 + i {\cal I}m(\int_0^\infty du \langle \Omega^*(0)\Omega(-u)\rangle e^{-i\nu u})$,
where
the function $\Gamma(E)$ is defined in Eq.~(3) of the main text.  
We then use the fact that   
 $\delta t$ is small compared to the evolution time of $\tilde \rho$, such that
$d\tilde\rho/dt= (\tilde{\rho} (t + \delta t) - \tilde{\rho}(t))/\delta t$.
Finally, Eq.~\eqref{eq:rhotildedeltatBM} gives
\begin{equation}
\begin{split}
  \frac{\mathrm{d}}{\mathrm{d}t}  \tilde{\rho} (t) = - \sum_{\mathbf{p}}  \left \{ \right . & \Gamma (\frac{\mathbf{p}^{2}}{2m} - g n - \epsilon_{\mathbf{p}}) u_{\mathbf{p}}^{2} ( \frac{1}{2} \{ a_{\mathbf{p}}^{\dag} a_{\mathbf{p}}, \tilde{\rho}(t) \} - a_{\mathbf{p}} \tilde{\rho}(t) a_{\mathbf{p}}^{\dag}) \\
      & +  \Gamma (\frac{ \mathbf{p}^{2}}{2m} - g n + \epsilon_{\mathbf{p}})  v_{\mathbf{p}}^{2} (\frac{1}{2} \{  a_{\mathbf{p}} a_{\mathbf{p}}^{\dag}, \tilde{\rho}(t) \} - a_{\mathbf{p}}^{\dag} \tilde{\rho}(t) a_{\mathbf{p}})\\
 &    - i  \left .\left [ \Delta_p + \Delta_p'a^+_{\bf p}a_{\bf p}, \tilde \rho \right ] \right \}
    \label{AppC11}
\end{split}
\end{equation}
where the commutator term, which corresponds to the effect of an hermitian term,
comes from the integrals in ${\cal I}m(\int_0^\infty du \langle \Omega^*(0)\Omega(-u)\rangle e^{-i\nu u})$
and amounts to the Lamb shift effect. It corresponds to a small correction of the
energies of the Bogoliubov modes, that we neglect in the following. 
Going back to the Shr\"odinger picture, we recover the master equation
given in Eq.~(12) of the main text. 

Eq.~(12) of the main text takes the form of a Lindblad equation. 
The originality in the above derivation is that the Born-Markov approximation can be done only
{\it after } the rotating wave approximation has been used to eliminate non resonant terms
corresponding to the term \eqref{eq:A2}. The Born-Markov approximation cannot be done before such
an approximation since we might consider Bogoliubov modes whose frequencies $\omega_{\bf p}$ can
be as large or higher than the reservoir frequency width $1/\tau_c$.

In the above  derivation of the master equation, we used a model of losses based on
a time-dependent noisy function. This permits to have simple expressions.
One could reproduce such calculations using instead a time-independent coupling
towards a continuum of state, such as the model considered in the appendix B. 

\section{Adiabatic following of $\langle a^+_{\bf p}a_{\bf p}\rangle$}
The Bogoliubov transformation given Eq.~(10) of the main text invert into
\begin{equation}
  \left \{
  \begin{array}{l}
    a_{\bf p}=u_p \Psi_{\bf p} -v_p \Psi^+_{-\bf p}\\
    a^+_{\bf p}=u_p \Psi^+_{\bf p} -v_p \Psi_{-\bf p}\\
    \end{array}
  \right .
  \end{equation}
The decrease of $n$ due to the loss process makes the functions $u_p$ and $v_p$ time dependent
and we note $\dot{u_p}=du_p/dt$ and $\dot{v_p}=dv_p/dt$.
Thus the operator $a^+_{\bf p}a_{\bf p}$ depends on time. Computing $d(a^+_{\bf p}a_{\bf p})/dt$
using the Bogoliubov transforms, and using the fact that, since 
$u_p^2-v_p^2=1$ for any time, $\dot{u}_pu_p -\dot{v_p}v_p=0$, we obtain
\begin{equation}
  \langle d (a^+_{\bf p}a_{\bf p})/dt \rangle=
 ( \dot{u}_p v_p - \dot{v}_p u_p )\left (
  \langle a_{-{\bf p}}a_{\bf p}\rangle +\langle a^+_{-{\bf p}}a^+_{\bf p}\rangle \right )
  \end{equation}
The mean values on the right hand side
typically oscillate at the frequency $2\epsilon_p/\hbar$. On the other hand we assume
slow losses such that
the evolution rate of $u_p$ and $v_p$ are much smaller than $\epsilon_p/\hbar$, and
the evolution  of $\langle a^+_{\bf p}a_{\bf p} \rangle$ is also expected to occur
on time scales much longer than $\hbar/\epsilon_p$.
Making a coarse grained approximation in time domain, the above equation then reduces to
\begin{equation}
  \langle d (a^+_{\bf p}a_{\bf p})/dt \rangle\simeq 0.
  \end{equation}
This approximation is a rotating wave approximation similar to that used in the
above appendix to derive the master equation. 

\section{Time evolution of the
temperature of the system}

We assume slow enough losses so that the gas has time to relax, at each time, towards a thermal
state, parameterized by the density $n$ and the temperature $T$.
The energy of the gas at time $t$ is $E(t)=E_{\rm th}(n(t),T(t))$ where
$E_{\rm th}(n,T)$, evaluated using the Bogoliuobv expression given in Eq.~(9) of the main text,
is
\begin{equation}
  E_{\rm th}(n,T)=e_0L^d +\sum_{{\bf p}\neq {\bf 0}} \epsilon_{p} n_{B}(\epsilon_{\bf p},T)
\end{equation}
where
\begin{equation}
  n_B(\epsilon,T)=\left ( e^{\epsilon/T}-1\right )^{-1}
\label{eq:thermalocc}
\end{equation}
is the Bose occupation factor of a mode of energy $\epsilon$.
The $n$-dependence of $E_{\rm th}$ comes from the $n$-dependence of $e_0$ and $\epsilon_{p}$, and
its dependence on $T$ comes from the $T$-dependence of the Bose occupation factors.

The density evolves as $dn/dt=-\Gamma_0 n$. This expression
is exact when the time correlation of the
reservoir is vanishing, and it is a very good approximation for reservoir of finite energy
width as long as the energy width is large enough.
One is then left in computing the time evolution of the temperature. 
The time derivative of $T$  is linked to  $dE/dt$ as
\begin{equation}
  \frac{dE}{dt}=\left (\frac{\partial E_{\rm th}}{\partial n}\right )_{T}(-\Gamma_0 n) + \left (
  \frac{\partial E_{\rm th}}{\partial T}\right )_{n} \frac{dT}{dt},
\end{equation}
which gives
\begin{equation}
  \begin{split}
  \frac{dE}{dt}=&-\Gamma_0 n L^d\frac{de_0}{dn} 
  -\Gamma_0 n \sum_{\bf p}  \frac{d\epsilon_p}{dn} \left ( n(\epsilon_p,T) +\epsilon_p \frac{\partial n_B(\epsilon_p,T)}{\partial \epsilon_p}\right )\\
  &+ \frac{dT}{dt}\sum_{\bf p} \epsilon_p \frac{dn_B(\epsilon_p,T)}{dT}
  \end{split}
  \label{eq:dEdtthermal}
\end{equation}

On the other hand, in the main text,  $dE/dt$ is computed
within the Bogoliubov theory and
using the master equation describing the effect of losses. The result is given by  Eq.~(13)
and Eq.~(14)  of the main text, where one should evaluate Eq.~(14) using for
$\langle a^+_pa_p\rangle$ the thermal
occupation factors given in Eq.~\eqref{eq:thermalocc}.
Comparing with Eq.~\eqref{eq:dEdtthermal}, and calculating explicitly
the partial derivatives of $n_B$, we find that $dT/dt$ fulfills
\begin{equation}
  	 \frac{dT}{dt}\sum_{\bf p}\frac{\epsilon_{\bf p}^2}{k_BT^2} \frac{e^{\epsilon_{\bf p}/k_BT}}{(e^{\epsilon_{\bf p}/k_BT} - 1)^2} = \sum_{\bf p}\epsilon_{\bf p}\left \{ \left (\frac{d \langle a_{\bf p}^+a_{\bf p}\rangle}{dt}\right )_{BG} 
	   - \Gamma_0 n 
\frac{\epsilon_p}{(k_B T)^2}\frac{e^{\epsilon_p/T}}{(e^{\epsilon_p/T}-1)^2}
         \frac{\partial \epsilon_p}{\partial n} \right \}
\end{equation}
where $\left ({d \langle a_{\bf p}^+a_{\bf p}\rangle}/{dt}\right )_{BG}$ is evaluated from Eq.~(14) of the main text, injecting the thermal Bose occupation factors $\langle a^+_{\bf p}a_{\bf p}\rangle=n_B(\epsilon_p,T)$.

We can apply this result in the two cases considered in the article.
In the case of contact interactions but a reservoir of finite energy width,
$dT/dt$ is computed by injecting 
$d\epsilon_p/dn= g p^2/(2m)/\epsilon_p$.
In the case of a reservoir of infinite energy width but for finite range interactions,
$dT/dt$ is computed by injecting 
$d\epsilon_p/dn= ge^{-p^2\sigma^2/2} p^2/(2m)/\epsilon_p$. Note also that in this case,
Eq.~(14) of the main text simplifies to
$\left ({d \langle a_{\bf p}^+a_{\bf p}\rangle}/{dt}\right )_{BG}= -\Gamma_0 (n_B(\epsilon_p,T)
+v_p^2)$.

\end{document}